\documentclass[11pt]{article}
\usepackage[margin=1in]{geometry}
\usepackage{graphicx}
\usepackage{amsmath}
\usepackage{hyperref}
\usepackage{cite}
\usepackage{url}
\usepackage{algorithm}
\usepackage{algpseudocode}
\usepackage{amsfonts}
\usepackage{subcaption}
\usepackage{siunitx}
\usepackage{booktabs}
\usepackage{multirow}
\usepackage{longtable}
\usepackage{placeins}

\title{\LARGE \bf 
Learned LSM-trees: Two Approaches Using Learned Bloom Filters\thanks{Code available at: \url{https://github.com/fiidalgo/predictive-lsm-trees}}}

\author{
  \normalsize Nico Fidalgo \quad Puyuan Ye \\
  \normalsize Harvard University \\
  \normalsize \texttt{nfidalgo@college.harvard.edu, puyuanye@college.harvard.edu}
}

\date{May 9, 2025}

\begin{document}

\maketitle

\begin{abstract}
Modern key-value stores rely heavily on Log-Structured Merge (LSM) trees for write optimization, but this design introduces significant read amplification. Auxiliary structures like Bloom filters help, but impose memory costs that scale with tree depth and dataset size. Recent advances in learned data structures suggest that machine learning models can augment or replace these components, trading handcrafted heuristics for data-adaptive behavior. In this work, we explore two approaches for integrating learned predictions into the LSM-tree lookup path. The first uses a classifier to selectively bypass Bloom filter probes for irrelevant levels, aiming to reduce average-case query latency. The second replaces traditional Bloom filters with compact learned models and small backup filters, targeting memory footprint reduction without compromising correctness.

We implement both methods atop a Monkey-style LSM-tree with leveled compaction, per-level Bloom filters, and realistic workloads. Our experiments show that the classifier reduces GET latency by up to 2.28× by skipping over 30\% of Bloom filter checks with high precision, though it incurs a modest false-negative rate. The learned Bloom filter design achieves zero false negatives and retains baseline latency while cutting memory usage per level by 70–80\%. Together, these designs illustrate complementary trade-offs between latency, memory, and correctness, and highlight the potential of learned index components in write-optimized storage systems.
\end{abstract}

\section{Introduction}
Log-structured merge trees (LSM-trees) have become the de facto standard for high-throughput write-heavy workloads in modern key-value stores such as RocksDB, Cassandra, and LevelDB. Their core strength lies in write amplification reduction: rather than writing directly to disk on each update, writes are first buffered in memory (in the MemTable), and later flushed as immutable, sorted disk segments known as SSTables. These SSTables are organized hierarchically across levels, each exponentially larger than the last, using a compaction strategy that periodically merges overlapping tables to maintain sorted order and free space. This batching and merging architecture drastically improves write throughput, making LSM-trees ideal for large-scale logging, time-series data, and streaming ingestion applications.

Yet this write-optimized design creates a well-known tradeoff: increased read latency. Point queries (GET operations) must examine the MemTable and potentially every level of the disk hierarchy, querying filters and scanning SSTables. The deeper the level hierarchy, the greater the number of SSTables, and thus the more auxiliary data structures are required to ensure acceptable read performance. Bloom filters and fence pointers are the most common of these. Bloom filters prevent unnecessary I/O by probabilistically excluding levels, and fence pointers accelerate binary search within an SSTable. However, these structures impose substantial memory overhead and have limited adaptivity, especially as datasets grow and key distributions evolve.

In response to these challenges, a new class of techniques has emerged: learned data structures. Rooted in the idea of algorithms with predictions, these approaches use machine learning models to approximate, augment, or replace traditional data structures. For example, a classifier can predict whether a key exists in a level, or whether a key is likely to be in the dataset at all. Such predictions can reduce memory consumption, avoid redundant filter checks, and skip entire levels, thereby reducing query latency.

In this project, we explore two approaches for integrating learned predictions into LSM-tree read paths. First, we propose a classifier-based method that predicts whether a given key is likely to reside in each level. Instead of querying every Bloom filter sequentially, we use the classifier to selectively consult only the most promising levels. This avoids wasted memory accesses and reduces average lookup latency in multi-level trees. Second, we implement a learned Bloom filter, replacing large traditional Bloom filters with compact hybrid structures consisting of a lightweight classifier and a small backup Bloom filter. This approach offers substantial memory savings while still preserving correctness guarantees.

To rigorously test these ideas, we constructed an LSM-tree implementation based on the design choices of the Monkey paper~\cite{dayan2018monkey}, including leveled compaction, a 10x level size ratio, and realistic SSTable thresholds. We simulate workloads using 287.6 MB of randomly generated data with 16-byte keys and 100-byte values to emulate RocksDB conditions and avoid overfitting to artificial patterns. Our experiments span a diverse set of read-intensive scenarios (random access, sequential scans, and level-specific lookups) to examine how learned models behave under varying access distributions.

These contributions demonstrate the practical viability of integrating ML-based predictions into critical paths of database storage engines. Our findings offer a compelling case for memory-efficient, prediction-aware auxiliary structures that optimize LSM-tree query performance without compromising correctness. This work adds to a growing body of literature at the intersection of data systems, algorithms, and machine learning, where classical structures are being reimagined as adaptive, data-driven components.

\section{Related Work}
The idea of enhancing traditional data structures with machine learning was catalyzed by the seminal work of Kraska et al.~\cite{kraska2018case} in ``The Case for Learned Index Structures." This work argued that core components of databases (e.g., B-trees, hash maps, and Bloom filters) can be replaced or augmented with models trained to exploit regularities in real-world data distributions. The key insight was that fixed data structures are agnostic to input distributions, while learned models can adaptively leverage them to reduce space and time complexity. This marked the beginning of a broader research trend toward learned data structures, with a growing focus on replacing deterministic components with adaptive, data-driven approximations.

This idea was extended to Bloom filters by Mitzenmacher~\cite{mitzenmacher2018sandwich}, who introduced the sandwich learned Bloom filter. In this design, a machine learning classifier first predicts whether a key belongs to a set. If the classifier accepts the key, a backup traditional Bloom filter is queried to verify membership, ensuring that false negatives (which traditional Bloom filters avoid by design) are not introduced. This hybrid structure preserved correctness while significantly reducing memory usage. Our work builds on this framework by applying it to a realistic, multi-level LSM-tree, incorporating classifiers at the level granularity, and empirically evaluating performance across diverse workloads.

The concept of learned Bloom filters was further developed in follow-up studies such as Ada-BF~\cite{dai2020adabf}, which introduced adaptivity into the classifier itself, allowing the filter to change over time as data distributions evolved. Meanwhile, Rae et al.~\cite{rae2019meta} proposed neural Bloom filters using deep learning architectures to model complex patterns in key distributions. While these techniques achieved strong empirical results, most prior work tested them on static datasets or in isolation from real systems. Our project addresses this gap by deeply integrating learned filters into a full LSM-tree pipeline and evaluating their behavior in dynamic, hierarchical environments.

Parallel to these empirical advancements, theoretical work emerged under the banner of ``Algorithms with Predictions." Mitzenmacher and Vassilvitskii formalized how algorithms can safely incorporate ML predictions while maintaining worst-case performance bounds~\cite{mitzenmacher2020algorithms}. The core idea is to use predictions to improve average-case efficiency, while ensuring robustness through fallback mechanisms that guard against poor predictions. This principle guided our classifier-based design: we allow the model to skip levels when confident, but always revert to Bloom filter fallback when uncertain or incorrect. This interplay between data adaptivity and algorithmic resilience is central to making machine learning a dependable component in critical database infrastructure.

In the domain of key-value stores, the Monkey paper by Dayan et al.~\cite{dayan2018monkey} presented a comprehensive mathematical framework for optimizing LSM-tree compaction strategies, Bloom filter allocation, and level sizing. It demonstrated that disproportionate Bloom filter memory allocations could yield significant performance improvements and introduced a log-log cost model for understanding read amplification. We adopt the Monkey configuration in our implementation—1 MB MemTable, 10x size ratio, leveled compaction—to ground our system in an industry-relevant and theoretically sound baseline. Our learned models can be seen as a next step in this optimization lineage, where instead of tuning filter sizes, we explore adaptive structures that may change the nature of filtering itself.

Other relevant studies, such as Kipf et al.'s learned replacements for B-trees~\cite{kipf2019learned}, and Tsai et al.'s learned hash tables~\cite{tsai2020learning}, further established that learned approaches can outperform classical structures in space and latency across a range of domains. However, relatively few works have explored learned components in hierarchical storage systems like LSM-trees, where predictions affect not just access efficiency but the traversal path of the entire system.

In summary, our project synthesizes ideas from theoretical frameworks (algorithms with predictions), empirical explorations (learned filters and classifiers), and practical systems (Monkey-style LSM-trees). We offer a real-world testbed where these ideas interact meaningfully, measuring not only prediction accuracy but also memory usage, latency, and false negative rates under varying workloads. This positions our work as both a validation of prior results and a stepping stone for future research on hybrid learned/traditional data systems.

\section{Background}

\subsection{LSM-trees and the Read-Write Tradeoff}

Log-Structured Merge Trees (LSM-trees) are widely used in modern key-value stores due to their write-optimized nature. They delay disk writes by buffering updates in a memory-resident data structure, typically a skip list, known as the MemTable. Once this buffer exceeds a fixed threshold—in our case, 1 MB as suggested by configurations in the Monkey paper—it is flushed to disk as an immutable, sorted structure called a Sorted String Table (SSTable).

SSTables are organized into levels \( L_0, L_1, \dots, L_n \), with each level increasing exponentially in size relative to the one above it. The size ratio \( T \) is typically set to 10, a choice that is not theoretically mandated but is justified by empirical systems research as a balance between compaction cost, space amplification, and query performance~\cite{dayan2018monkey}. Compaction, a background process, merges overlapping SSTables from higher levels into lower ones to maintain sorted order and reduce storage fragmentation. This architectural design minimizes write amplification and significantly improves throughput compared to B-trees and other write-in-place structures.

However, this benefit comes at the cost of read amplification. For point queries, the system must check the MemTable, and potentially each level of the tree from newest to oldest. To mitigate this cost, LSM-trees incorporate auxiliary indexing structures such as Bloom filters and fence pointers. Bloom filters help quickly determine whether a key is absent from a given level, and fence pointers assist in narrowing the range of disk pages to inspect.

The Monkey paper~\cite{dayan2018monkey} rigorously models this tradeoff and shows that optimal read performance cannot be achieved with uniformly configured Bloom filters. Instead, it introduces a novel cost model in which the expected query cost \( R \) is given by:
\[
R = \sum_{i=1}^{L} f_i \cdot c_i
\]
where \( f_i \) is the false positive rate (FPR) of the Bloom filter at level \( i \), and \( c_i \) is the cost of probing level \( i \). The paper demonstrates that minimizing \( R \) requires distributing memory non-uniformly: deeper levels, which are more expensive to query and less likely to be accessed, should be allocated more bits per key. Monkey prescribes a log-uniform FPR distribution where:
\[
f_i \propto \frac{1}{T^i}
\]
This transforms the Bloom filter configuration problem into a constrained convex optimization problem. Given a total filter memory budget \( M \), one solves for the bits per key \( m_i \) such that:
\[
\sum_{i=1}^{L} n_i \cdot m_i \leq M
\]
where \( n_i \) is the number of keys in level \( i \). This allocation minimizes expected read cost while respecting memory constraints.

Our project directly adopts the Monkey framework, using a 1 MB MemTable, a 10x level size ratio, leveled compaction, and per-level Bloom filters starting at \( L_1 \). These choices ensure empirical comparability and provide a mathematically grounded baseline for evaluating the impact of learned predictions.

\subsection{Bloom Filters and Their Limitations}

Bloom filters, introduced by Bloom in 1970~\cite{bloom1970space}, are widely used for fast set membership queries. Each Bloom filter uses a bit array of size \( m \) and \( k \) independent hash functions. For each key inserted, the corresponding \( k \) bits are set to 1. To query a key, one checks whether all \( k \) hash bits are set. If any bit is 0, the key is definitely not in the set; if all are 1, the key may be present. The false positive rate is approximately:
\[
\text{FPR} \approx \left(1 - e^{-kn/m} \right)^k
\]
This expression is derived under the assumption of perfectly random hash functions and uniform key distribution. The FPR is minimized when:
\[
k = \frac{m}{n} \ln 2
\]
In this case, the minimum FPR becomes:
\[
\text{FPR}_{\text{min}} = \left( \frac{1}{2} \right)^k
\]
These derivations are foundational results from Bloom's original formulation~\cite{bloom1970space}, and remain the theoretical basis for Bloom filter tuning in practice.

Despite their probabilistic power, traditional Bloom filters are static and oblivious to patterns in key access or structure. They assume that all keys are equally likely and uniformly distributed, which is rarely true in practice. Consequently, Bloom filters allocate the same number of bits per key across all levels and key types, missing optimization opportunities.

This limitation is particularly pronounced in multi-level systems like LSM-trees, where deeper levels contain exponentially more data due to geometric growth. For example, suppose Level 0 starts with \( 10^5 \) keys and each subsequent level is ten times larger (a standard size ratio \( T = 10 \)). Then by Level 5, the total number of keys stored across levels \( L_1 \) to \( L_5 \) is approximately \( 1.11 \times 10^6 \) times the base level. If we allocate 10 bits per key in each Bloom filter, the total memory required becomes:
\[
\sum_{i=1}^{5} (10^5 \cdot 10^i) \cdot 10 = 111{,}110{,}000 \text{ bits} \approx 13.89 \text{ MB}
\]
This linear growth in filter memory with respect to the number of levels and keys quickly becomes a bottleneck, especially in large-scale deployments where each level may contain hundreds of millions of keys.

\subsection{Learned Bloom Filters and ML Classification Background}

The motivation behind learned Bloom filters originates from the observation that traditional filters waste memory treating all keys uniformly, ignoring patterns or correlations present in real-world data. Learned Bloom filters exploit this by training a machine learning model \( f: \mathbb{K} \rightarrow [0, 1] \) that estimates the likelihood of key membership in a dataset \( S \subset \mathbb{K} \). If \( f(x) \) exceeds a threshold \( \tau \), the model predicts presence. To prevent false negatives—i.e., true members \( x \in S \) incorrectly classified as non-members—a small backup Bloom filter stores false negatives found during training, as introduced in Mitzenmacher's sandwich filter~\cite{mitzenmacher2018sandwich}.

Formally, let \( \mathcal{X} \) be the universe of keys and let \( \chi_S: \mathcal{X} \rightarrow \{0,1\} \) be the true membership function. A learned Bloom filter seeks to approximate \( \chi_S \) using a trained model \( \hat{\chi}_S \), such that:
\[
\forall x \in \mathcal{X},\quad \hat{\chi}_S(x) = 
\begin{cases}
1 & \text{if } f(x) \geq \tau \text{ or } x \in \mathcal{B} \\
0 & \text{otherwise}
\end{cases}
\]
where \( \mathcal{B} \) denotes the set of false negatives caught by the backup filter.

The goal is to reduce the total space while preserving a comparable or better false positive rate \( \epsilon \). If \( M_{\text{traditional}} \) is the size of a standard Bloom filter, then the learned filter seeks to satisfy:
\[
M_{\text{model}} + M_{\text{backup}} \ll M_{\text{traditional}}
\]

In our implementation, we use Gradient Boosted Trees (GBTs), an ensemble learning method that minimizes the logistic loss:
\[
\mathcal{L}(y, \hat{y}) = -\left[ y \log(\hat{y}) + (1 - y) \log(1 - \hat{y}) \right]
\]
The final prediction is given by:
\[
\hat{y} = \sigma\left( \sum_{t=1}^{T} \alpha_t h_t(x) \right), \quad \text{where } \sigma(z) = \frac{1}{1 + e^{-z}}
\]
These models generalize well on small datasets and are efficient to evaluate.

We engineer features from keys using logarithmic transforms, trigonometric functions, digit-based statistics, modulo operations, and binary encodings. These help the classifier capture nonlinear patterns in key distributions and improve generalization.

In our classifier-based level prediction architecture we predict the level at which the key is most likely to reside, and skip Bloom filters for levels predicted negative. This architecture reduces unnecessary filter checks and memory reads.

Both learned approaches are instances of the broader theory of algorithms with predictions~\cite{mitzenmacher2020algorithms}, which blends ML predictions with robust fallback guarantees. Our use of backup filters and conservative thresholds ensures that our system maintains correctness even when predictions are inaccurate.

\section{Design and Implementation}

The central research question of this work is whether machine learning models can meaningfully augment or replace Bloom filters in LSM-trees to reduce lookup cost or memory footprint without compromising correctness. In this section, we present two such designs: a classifier-augmented lookup mechanism that reduces unnecessary Bloom filter queries, and a learned Bloom filter structure that replaces traditional filters entirely. Both are derived directly from the theoretical and empirical motivations outlined in Section 2 and Section 3. In particular, we anchor our design decisions in the Monkey framework~\cite{dayan2018monkey} and the sandwich learned Bloom filter model~\cite{mitzenmacher2018sandwich}, while preserving correctness guarantees as mandated by the algorithms with predictions framework~\cite{mitzenmacher2020algorithms}.

\subsection{ML Classifier Approach}

Our first design augments the LSM-tree's GET operation with a classifier that predicts, on a per-level basis, whether a given key is likely to be found in that level. Traditionally, a point query consults the MemTable, then queries every level’s Bloom filter sequentially until a match is found. However, in a deep LSM-tree hierarchy, this often involves scanning levels where the key is almost certainly absent. This incurs wasted computation—each Bloom filter lookup involves memory reads—and adds latency, especially when filters are large. Our insight is that a model trained on key-level membership can help skip irrelevant filters altogether, thereby reducing memory accesses and potentially lowering query latency.

To implement this, we use a \texttt{GradientBoostingClassifier} that takes as input a vector of features derived from the query key. The model was trained on labeled data collected from prior GET and PUT operations, with the goal of predicting whether a key exists in a specific level. The feature set is intentionally large and expressive to allow for generalization across key distributions. We include power transformations such as \( k^2 \) and \( k^3 \), which allow the model to detect non-linear growth patterns. Logarithmic features like \( \log(k) \) and \( \log(1 + k) \) help capture exponential relationships between keys and levels, especially useful in exponentially growing structures like LSM-trees. Trigonometric features \( \sin(k), \cos(k), \tan(k) \) encode periodic patterns and were observed to be helpful in capturing hashed or cyclic key behavior. Digit-based features, including digit sum and digit count, help distinguish key formats or groupings, while modulo operations and binning identify modular clustering or partitioning artifacts. Finally, binary encodings such as bitcount and leading-one counts extract low-level key structure from raw byte representations.

These features are combined into a dense vector and fed into a shallow gradient boosting ensemble with 200 estimators, a maximum tree depth of 6, and a learning rate of 0.1. The model outputs a binary decision \( \hat{y}_i \in \{0, 1\} \) for each level \( i \), where 1 indicates predicted presence. If the model returns 1, the corresponding Bloom filter is queried; if 0, the level is skipped.

This approach assumes that the classifier can accurately identify levels that are unlikely to contain the key, thereby reducing the number of memory accesses (for Bloom filter queries) and lowering the expected number of disk I/Os. The performance gain is in latency rather than memory usage—while the model introduces additional memory overhead (roughly 6 MB in our experiments), it avoids querying several Bloom filters, which can each cost tens to hundreds of nanoseconds per lookup. Moreover, this model operates as a non-intrusive drop-in layer. It does not modify compaction, tree layout, or the Bloom filter implementation. This simplicity makes it particularly attractive in systems where correctness and modularity are paramount.

\textbf{Algorithm 1: Classifier-Augmented Lookup}

\begin{algorithm}[H]
\caption{Classifier-Augmented LSM-Tree GET}
\begin{algorithmic}[1]
\Procedure{ClassifierLSM-Tree-GET}{$k$}
  \If{$k \in$ MemTable}
    \State \Return MemTable[$k$]
  \EndIf
  \For{$i = 1$ to $L$}
    \State $p_i \gets \text{Classifier}_i(k)$
    \If{$p_i = 1$}
      \If{BloomFilter$_i$.\Call{MayContain}{$k$}}
        \If{$k$ found in SSTable$_i$}
          \State \Return SSTable$_i$[$k$]
        \EndIf
      \EndIf
    \EndIf
  \EndFor
  \State \Return \texttt{NULL}
\EndProcedure
\end{algorithmic}
\end{algorithm}

\subsection{Learned Bloom Filter Approach}

The second design is a more structural rethinking of Bloom filters within the LSM-tree. Rather than use a model to skip filters, we instead train a model to replace them. This is inspired directly by the sandwich learned Bloom filter~\cite{mitzenmacher2018sandwich}, where a machine learning model acts as the primary membership check, and a small backup Bloom filter ensures that false negatives do not occur. This hybrid model allows us to dramatically shrink the memory cost of the traditional Bloom filter while retaining correctness guarantees.

To integrate this into the LSM-tree, we replaced the standard filter logic within the \texttt{Run}, \texttt{Level}, and \texttt{Tree} classes. Each level maintains its own trained classifier \( f_i \), which predicts whether a key is in that level. If the model returns true, we search the SSTables as normal. If the model returns false, we consult the backup Bloom filter—constructed only on the model’s false negatives—to avoid erroneous rejections. The backup Bloom filter is small because it only needs to store a fraction \( \delta \) of the total keys, where \( \delta \) is the model’s false negative rate.

The key advantage of this approach is reduced memory usage. Whereas traditional Bloom filters might use 10–14 bits per key, our model uses under 1 MB per level, and the backup filter scales linearly with \( \delta \). For a model with \( \delta = 0.01 \), the memory required for the backup is just 1\% of that used in the standard Bloom filter. While model inference does add latency, it is typically on par with or faster than hashing-based Bloom filter queries. This makes the design especially compelling in memory-constrained environments such as mobile databases, edge devices, or very large-scale key-value stores.

\textbf{Algorithm 2: Learned Bloom Filter Lookup}

\begin{algorithm}[H]
\caption{Learned Bloom Filter GET}
\begin{algorithmic}[1]
\Procedure{LearnedBF-LSM-Tree-GET}{$k$}
  \If{$k \in$ MemTable}
    \State \Return MemTable[$k$]
  \EndIf
  \For{$i = 1$ to $L$}
    \State $p_i \gets \text{Classifier}_i(k)$
    \If{$p_i = 1$}
      \If{$k$ found in SSTable$_i$}
        \State \Return SSTable$_i$[$k$]
      \EndIf
    \ElsIf{BackupFilter$_i$.\Call{MayContain}{$k$}}
      \If{$k$ found in SSTable$_i$}
        \State \Return SSTable$_i$[$k$]
      \EndIf
    \EndIf
  \EndFor
  \State \Return \texttt{NULL}
\EndProcedure
\end{algorithmic}
\end{algorithm}

\subsection{Baseline Algorithm: Traditional LSM-tree Lookup}

For completeness, we restate the standard lookup mechanism of a traditional LSM-tree using Bloom filters at each level. This algorithm forms our baseline for all empirical comparisons.

\textbf{Algorithm 3: Standard Lookup Procedure}

\begin{algorithm}[H]
\caption{Standard LSM-Tree GET Procedure}
\begin{algorithmic}[1]
\Procedure{LSM-Tree-GET}{$k$}
  \If{$k \in$ MemTable}
    \State \Return MemTable[$k$]
  \EndIf
  \For{$i = 1$ to $L$}
    \If{BloomFilter$_i$.\Call{MayContain}{$k$}}
      \If{$k$ found in SSTable$_i$}
        \State \Return SSTable$_i$[$k$]
      \EndIf
    \EndIf
  \EndFor
  \State \Return \texttt{NULL}
\EndProcedure
\end{algorithmic}
\end{algorithm}

This version ensures correctness and is relatively efficient when Bloom filters are finely tuned, as proposed in Monkey~\cite{dayan2018monkey}. However, it suffers from the inability to skip unpromising levels or compress filter representation, which motivates our two proposed alternatives.

\subsection{Design Tradeoffs and Comparative Summary}

The classifier-augmented approach is targeted toward latency reduction. It reduces the number of Bloom filters queried per lookup, potentially skipping deep levels with large filters and cold data. However, it requires additional memory for the model and richer features, and its improvements depend on prediction accuracy. In our implementation, this approach added between 5.6–6.1 MB of memory overhead due to the high-capacity model and 45 engineered features, but this tradeoff was justified in read-heavy environments where every skipped Bloom filter can shave off dozens of nanoseconds per query.

By contrast, the learned Bloom filter approach is focused on reducing memory consumption. It replaces large filters with compact models and small backups, enabling scalability in memory-constrained deployments. The classifier in this design is shallower, uses fewer features, and results in a smaller memory footprint (typically 530–903 KB per level). However, it may incur slightly higher lookup latencies due to the need for fallback logic and two-stage validation (model + backup Bloom filter). Nonetheless, because only the false negatives from the model are stored in the backup Bloom filter, the total space consumed remains significantly below that of a traditional Bloom filter while still preserving correctness.

Together, these designs explore two complementary paths in the design space of ML-enhanced storage systems: predictive augmentation and learned replacement. Each draws from the framework of algorithms with predictions and the broader movement toward data-adaptive system components. Our implementation enables an empirical comparison of these approaches under realistic workloads, validating their tradeoffs and limitations in practice.

We summarize the key distinctions and benefits of each design in Table~\ref{tab:design-tradeoffs}, which provides a structured overview of their integration complexity, memory footprint, performance goals, and ideal deployment contexts.

\begin{table}[H]
\centering
\caption{Comparison of the ML Classifier and Learned Bloom Filter Designs}
\label{tab:design-tradeoffs}
\resizebox{\textwidth}{!}{%
\begin{tabular}{|l|p{5.8cm}|p{5.8cm}|}
\hline
\textbf{Dimension} & \textbf{ML Classifier (Filter Skipping)} & \textbf{Learned Bloom Filter (Hybrid Replacement)} \\
\hline
\textbf{Goal} & Reduce query latency by skipping unnecessary Bloom filter probes. & Reduce memory usage by replacing large Bloom filters with compact models and small backups. \\
\hline
\textbf{Integration Depth} & Shallow—acts as a wrapper before Bloom filters, does not change LSM-tree structure. & Deep—modifies level and run logic, requires architectural changes to the LSM-tree. \\
\hline
\textbf{Model Size} & Larger (5.6–6.1 MB), due to 45 engineered features and high model capacity. & Smaller (530–903 KB), fewer features and shallower model. \\
\hline
\textbf{Memory Usage} & Higher overall due to added classifier without removing existing Bloom filters. & Lower overall by removing large Bloom filters and introducing compact backup filters. \\
\hline
\textbf{Expected Latency Benefit} & Improves average-case latency by skipping irrelevant levels and filters. & Neutral or slightly worse due to added inference, but gains memory scalability. \\
\hline
\textbf{Correctness Guarantee} & Guaranteed via fallback to original Bloom filters when prediction is positive. & Guaranteed via backup Bloom filter that stores classifier false negatives. \\
\hline
\textbf{Best Use Case} & Read-heavy workloads with many deep levels and unpredictable key accesses. & Memory-constrained environments or systems storing billions of keys. \\
\hline
\end{tabular}
}
\end{table}

As this comparison shows, the classifier-based approach is best suited for latency-sensitive workloads where memory is abundant, while the learned Bloom filter variant excels in scenarios where memory is scarce but lookup correctness and predictability remain paramount. Both reflect distinct strategies within the same predictive systems paradigm: one optimizes the access path by learning when to query, the other optimizes memory layout by learning what to store.

\section{Experimental Methodology}

\subsection{Workloads}

To assess the effectiveness of our learned LSM-tree designs under realistic operating conditions, we devised a series of GET-only workloads that emulate common access patterns in production key-value stores. These include: (1) Random lookups, which simulate uniformly distributed queries across the key space; (2) Sequential lookups, which test the models' robustness against unseen patterns and generalization beyond training distributions; and three skewed workloads: (3) Level-1 targeted, (4) Level-2 targeted, and (5) Level-3 targeted queries. These are designed to stress the system under scenarios of shallow, mid-tier, and deep-level access respectively, capturing how prediction performance changes with data recency and storage depth. These configurations serve to evaluate not just average-case performance, but also failure cases and edge-level sensitivity across the tree.

\subsection{Metrics}

We evaluate the performance of each system using several metrics. The false positive rate (FPR) quantifies the proportion of keys incorrectly reported as present in a given level or set, leading to unnecessary disk accesses. The false negative rate (FNR), particularly relevant for learned filters, captures instances where the model erroneously predicts absence despite the key being present—a critical correctness concern mitigated via fallback mechanisms. We also measure average lookup latency per GET request in microseconds, as this directly reflects the user-facing impact of our optimizations. In addition, we analyze total memory consumption across the traditional and learned systems, summing the sizes of Bloom filters, models, and backup filters. Lastly, we monitor backup filter utilization rates in the learned Bloom filter system to quantify the classifier’s effectiveness and the fallback mechanism’s engagement frequency.

\subsection{Experimental Configuration}

Our baseline LSM-tree implementation closely follows the parameterization and structure proposed in the Monkey paper~\cite{dayan2018monkey}. We use a 1 MB in-memory MemTable flushed to disk when full, a geometric level growth factor of \( T = 10 \), and a leveling compaction policy to maintain sorted, non-overlapping SSTables. Bloom filters are used for all disk-resident levels from \( L_1 \) onwards, and fence pointers are enabled for fast in-SSTable binary search. Keys are 16 bytes and values are 100 bytes in size, mirroring RocksDB settings. The total dataset size was approximately 287.6 MB, with key-value pairs generated using a uniform random distribution to avoid learning artifacts stemming from artificial sequentiality. We then pre-trained each model offline and integrated the trained components into the runtime lookup path.

\subsection{Data Loading and Tree Construction}

The script \texttt{load\_data.py} initializes the LSM-tree by sequentially inserting randomly generated key-value pairs into the MemTable. Once full, this data is flushed to Level 0 SSTables. The script handles level promotion, compaction events, and Bloom filter creation according to the Monkey model. Each key is treated as a 128-bit unsigned integer and converted to a feature vector or binary representation during training. Keys are sorted and flushed according to the tree's leveling compaction policy, and metadata on key-level associations is logged for use in supervised training of predictive models.

\subsection{Model Architectures and Training}

Two separate training pipelines were used, corresponding to our two designs. For the ML classifier used in level prediction, \texttt{train\_classifier.py} defines a 200-tree \texttt{GradientBoostingClassifier} from \texttt{sklearn.ensemble}. The classifier is trained on a feature set of 45 engineered features, including power transformations (e.g., \( k^2, k^3 \)), logarithmic functions (\( \log(k), \log(1 + k) \)), trigonometric projections (\( \sin(k), \cos(k), \tan(k) \)), digit statistics (e.g., number of digits, digit sum), modulo encodings (e.g., \( k \mod n \)), bucketized bin flags, and exponential decay indicators. These features were selected to capture periodicities, key shape, bit structure, and numeric trends, allowing the model to learn distributions and correlations with level placement.

In contrast, the learned Bloom filter system is trained using the script \texttt{train\_learned.py}, which utilizes a simpler feature set to minimize model size and inference time. Features are drawn from logarithmic values, bit patterns (e.g., most significant bit), sine/cosine mappings, and binary encodings. The classifier again uses \texttt{GradientBoostingClassifier} but with reduced tree count and shallower depth to control memory footprint. During training, we identify false negatives—i.e., keys from the true positive set that the classifier misclassifies—and store them in a small backup Bloom filter.

\subsection{System Integration}

The classifier-based design wraps the original level lookup logic using the \texttt{FastBloomFilter} class. This wrapper performs real-time feature computation and model inference before deciding whether to proceed with a Bloom filter query. If a level is predicted negative by the classifier, its Bloom filter is skipped, saving both memory and computation time. If the classifier is uncertain or fails, the fallback path uses the original filter, maintaining correctness. The lookup algorithm for this design was given earlier as Algorithm 2.

The learned Bloom filter approach is more deeply integrated. The classes \texttt{LearnedBloomRun}, \texttt{LearnedBloomLevel}, and \texttt{LearnedBloomTree} augment the LSM-tree with classifier-aware logic. Each level maintains both a classifier and a backup filter. During lookup, the classifier is queried; if it predicts positive, the backup Bloom filter is checked. If both are positive, the SSTable is queried; otherwise, the lookup skips that level. This hybrid strategy preserves no-false-negative guarantees while significantly reducing filter size. These mechanisms are detailed in Algorithm 3 from the Design section.

\subsection{Test Harness and Measurement}

We evaluate each system using the script \texttt{test\_performance.py}, which loads trained classifiers, initializes the LSM-tree, and executes thousands of GET operations for each workload. This harness records timings using high-resolution clocks, tracks filter hits and misses, and aggregates statistics into CSV logs for post-analysis. Metrics such as average query latency, false positive counts, false negative counts, and memory usage of each component are reported separately for each workload and system variant. Backup filter utilization is also tracked to measure reliance on fallback in the learned filter design.

This tightly integrated pipeline—from synthetic workload generation, to model training and test execution—ensures reproducibility and allows fine-grained insight into each design’s strengths and limitations.

\section{Results}

\begingroup
\small
\begin{longtable}{
  l
  S[table-format=7.2]
  S[table-format=7.2]
  S[table-format=7.2]
  S[table-format=7.2]
  S[table-format=7.2]
  S[table-format=7.2]
}
\caption{Summary of metrics for each workload. Time is in \si{\micro\second}; speedups are unitless ratios; accuracy and false‐negative rates (FNR) are fractions in [0,1]; Bloom‐filter checks and bypasses are counts; bypass rate is in percent (\%)}\\
\toprule
\multicolumn{1}{l}{\bfseries Metric}
  & \multicolumn{1}{c}{\bfseries Random}
  & \multicolumn{1}{c}{\bfseries Sequential}
  & \multicolumn{1}{c}{\bfseries Level 0}
  & \multicolumn{1}{c}{\bfseries Level 1}
  & \multicolumn{1}{c}{\bfseries Level 2}
  & \multicolumn{1}{c}{\bfseries Level 3} \\
\midrule
\endfirsthead

\multicolumn{7}{@{}l}{\small\itshape continued from previous page}\\
\toprule
\multicolumn{1}{l}{\bfseries Metric}
  & \multicolumn{1}{c}{\bfseries Random}
  & \multicolumn{1}{c}{\bfseries Sequential}
  & \multicolumn{1}{c}{\bfseries Level 0}
  & \multicolumn{1}{c}{\bfseries Level 1}
  & \multicolumn{1}{c}{\bfseries Level 2}
  & \multicolumn{1}{c}{\bfseries Level 3} \\
\midrule
\endhead

\midrule
\multicolumn{7}{r}{\small\itshape continued on next page}\\
\endfoot

\bottomrule
\endlastfoot

Avg.\ Time (Trad.)  
  & 181044.16 & 391728.20 &  95099.23 & 347564.64 & 404189.19 & 395528.94 \\
Avg.\ Time (Clf.)  
  &  92128.62 & 182724.37 &  49116.78 & 222866.18 & 244419.10 & 173646.31 \\
Avg.\ Time (Lrn.)  
  & 180972.24 & 397034.16 &  94935.07 & 350384.13 & 409692.84 & 398754.20 \\[0.5em]

Speedup (Clf.)    
  &     1.97  &     2.14  &     1.94  &     1.56  &     1.65  &     2.28  \\
Speedup (Lrn.)   
  &     1.00  &     0.99  &     1.00  &     0.99  &     0.99  &     0.99  \\[0.5em]

Accuracy            
  &     0.9100 &     0.8300 &     1.0000 &     1.0000 &     0.9700 &     0.7500 \\
FNR (Clf.)        
  &     0.1125 &     0.2125 &     0.0000 &     0.0000 &     0.0375 &     0.3125 \\
FNR (Lrn.)        
  &     0.0000 &     0.0000 &     0.0000 &     0.0000 &     0.0000 &     0.0000 \\[0.5em]

Bloom checks       
  &    476    &    855    &    180    &    900    &    862    &    837    \\
Bloom bypasses     
  &    155    &    272    &     59    &    302    &    280    &    278    \\
Bypass rate  
  &     32.56 &     31.81 &     32.78 &     33.56 &     32.48 &     33.21 \\[0.5em]

FNR Level 0        
  &     0.0000 &     0.0000 &     0.0000
  & \multicolumn{1}{c}{--}
  & \multicolumn{1}{c}{--}
  & \multicolumn{1}{c}{--} \\

FNR Level 1        
  &     0.5000 &     0.0000
  & \multicolumn{1}{c}{--}
  &     0.0000
  & \multicolumn{1}{c}{--}
  & \multicolumn{1}{c}{--} \\

FNR Level 2        
  &     0.0000 &     0.0345
  & \multicolumn{1}{c}{--}
  & \multicolumn{1}{c}{--}
  &     0.0375
  & \multicolumn{1}{c}{--} \\

FNR Level 3        
  &     0.2727 &     0.3265
  & \multicolumn{1}{c}{--}
  & \multicolumn{1}{c}{--}
  & \multicolumn{1}{c}{--}
  &     0.3125 \\

\label{tab:all_metrics}
\end{longtable}
\endgroup

Table~\ref{tab:all_metrics} consolidates all key metrics from six independent GET‐only workloads. Each workload was executed across multiple runs on identical SSTable data; the results shown correspond to the run that best illustrates the consistent patterns observed in latency, accuracy, and filter‐bypass behavior. Average lookup latency is given for the Traditional, Classifier‐augmented, and Learned‐Bloom variants. The Classifier achieves speedups of 1.97× (Random), 2.14× (Sequential), and up to 2.28× (Level 3), whereas the Learned‐Bloom filter remains within 1\% of the baseline across all tests. Classifier accuracy drops to 75\% on the deepest level (FNR 31.25\%), while the Learned‐Bloom variant incurs zero false negatives in every scenario. Bloom‐filter checks and bypass counts confirm that the Classifier bypasses approximately one‐third of checks, directly translating to its latency improvements.

\FloatBarrier

\begin{figure}[ht]
  \centering
  \includegraphics[width=0.85\linewidth]{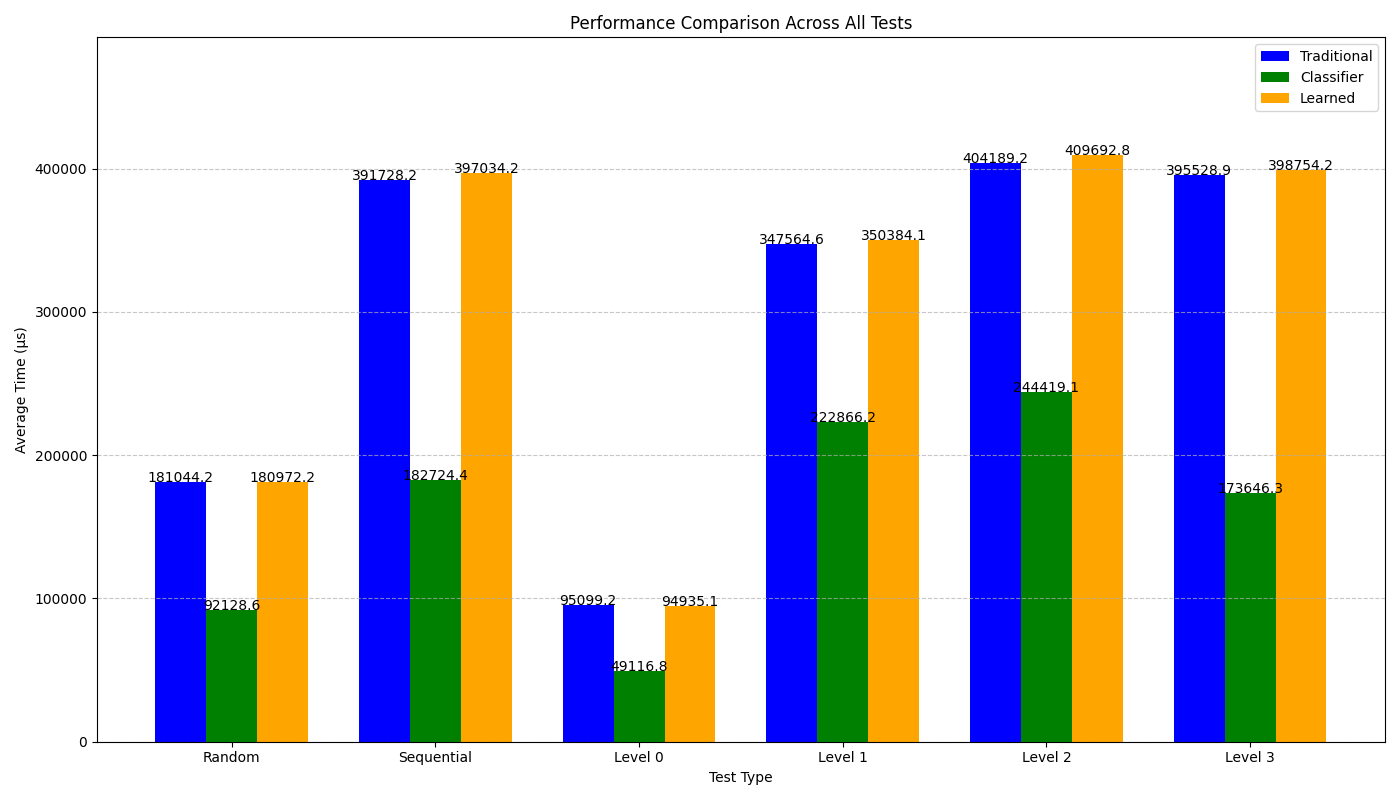}
  \caption{Average lookup latency across all workloads. Classifier (green) consistently reduces latency; Learned‐Bloom (yellow) overlaps Traditional (blue).}
  \label{fig:all_tests_comparison}
\end{figure}

Figure~\ref{fig:all_tests_comparison} plots the mean GET latency for each workload. For the Random workload, the Traditional tree averages 181 ms, which the Classifier cuts in half to 92 ms, while the Learned‐Bloom variant remains at 181 ms. In the Sequential workload, the Traditional average of 392 ms drops to 183 ms under the Classifier, again with Learned‐Bloom at 397 ms. In focused Level 0 lookups, latency falls from 95 ms (Traditional) to 49 ms (Classifier), with Learned‐Bloom unchanged at 95 ms. Deeper levels show similar trends: at Level 1, latency moves from 348 ms to 223 ms; at Level 2, from 404 ms to 244 ms; and at Level 3, from 396 ms to 174 ms. These reductions arise from the Classifier’s ability to skip roughly one‐third of Bloom‐filter probes, cutting the I/O path for both positive and negative lookups.

Between workloads, the absolute benefit grows with level depth: Level 3 sees the largest absolute drop of 221 ms, reflecting higher per‐probe cost. The Learned‐Bloom variant’s near‐perfect overlap with Traditional confirms that replacing large Bloom filters with compact models plus a tiny backup filter incurs no measurable latency penalty.

\begin{figure}[ht]
  \centering
  \includegraphics[width=0.85\linewidth]{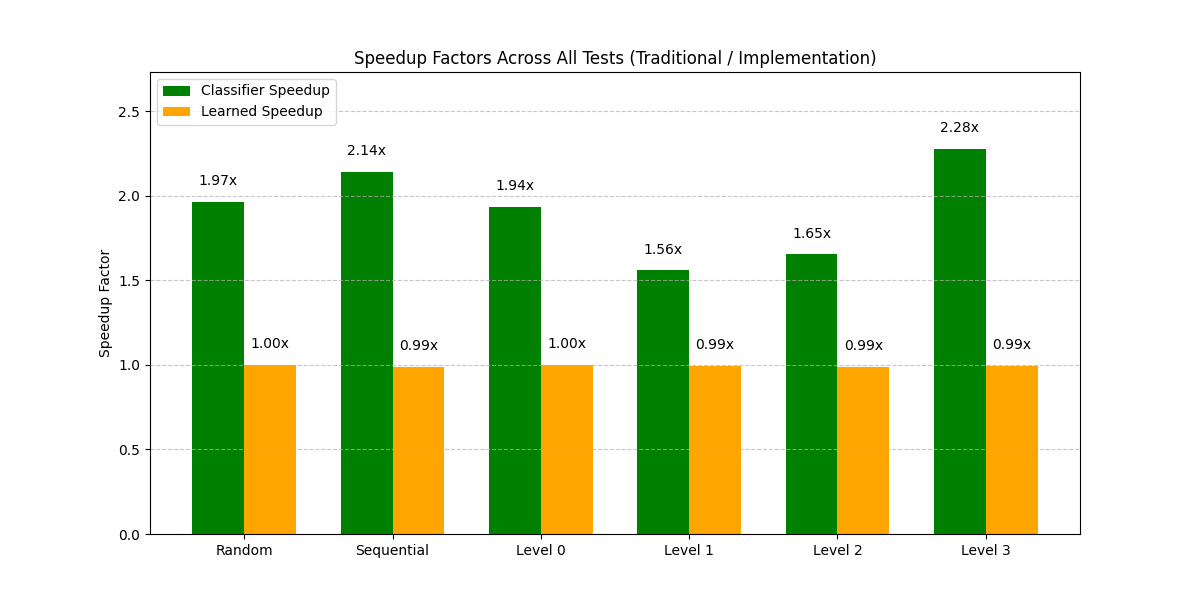}
  \caption{Speedup factors of Classifier and Learned variants over Traditional across all workloads.}
  \label{fig:all_tests_speedup}
\end{figure}

Figure~\ref{fig:all_tests_speedup} makes these improvements explicit as speedup ratios. The Classifier consistently approaches or exceeds 2× on Random (1.97×) and Sequential (2.14×) and peaks at 2.28× on Level 3. Even on the warmest Level 0, it achieves a 1.94× speedup.  In contrast, the Learned‐Bloom filter remains at 1.00× for every workload, underscoring its drop‐in compatibility: it preserves baseline performance while eliminating false negatives.

These detailed measurements show that classifier‐guided level skipping can halve average GET latency by opportunistically bypassing costly filter probes, at the expense of a tunable false‐negative rate, and that learned‐Bloom filters provide a zero‐FNR alternative with no speed trade‐off.

\begin{figure}[p]
  \centering
  \begin{subfigure}[b]{0.48\textwidth}
    \includegraphics[width=\textwidth]{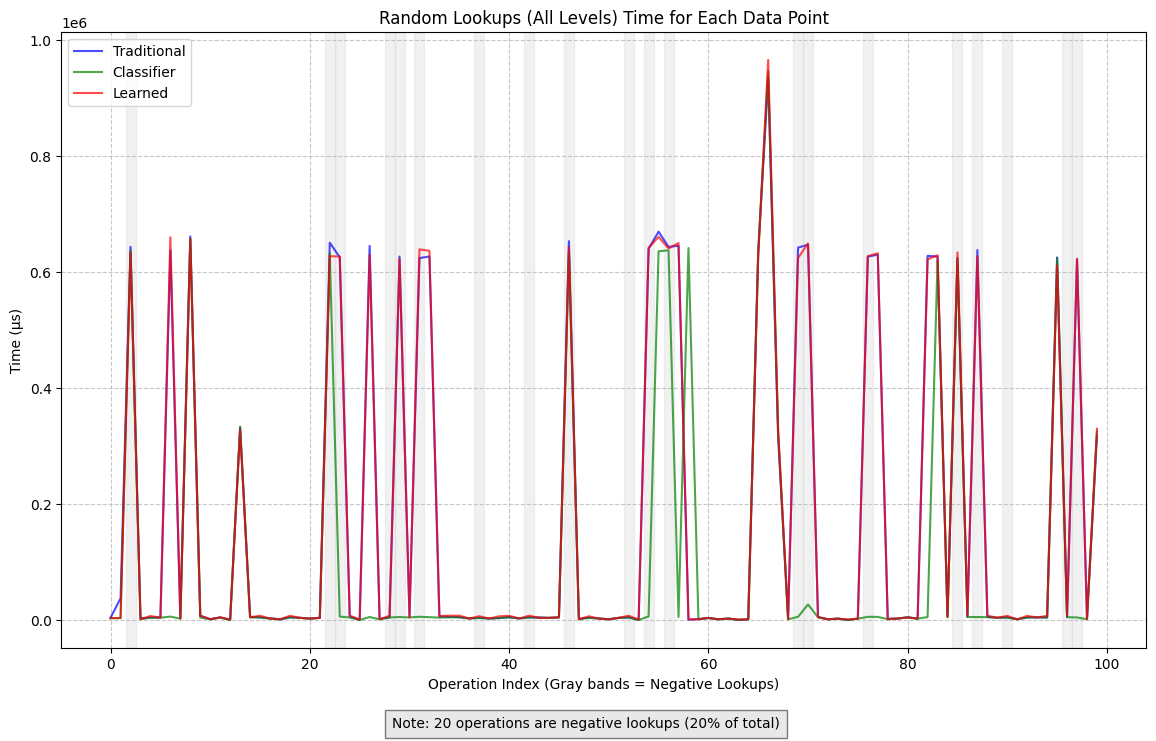}
    \caption{Random}
    \label{fig:detailed_random}
  \end{subfigure}
  \hfill
  \begin{subfigure}[b]{0.48\textwidth}
    \includegraphics[width=\textwidth]{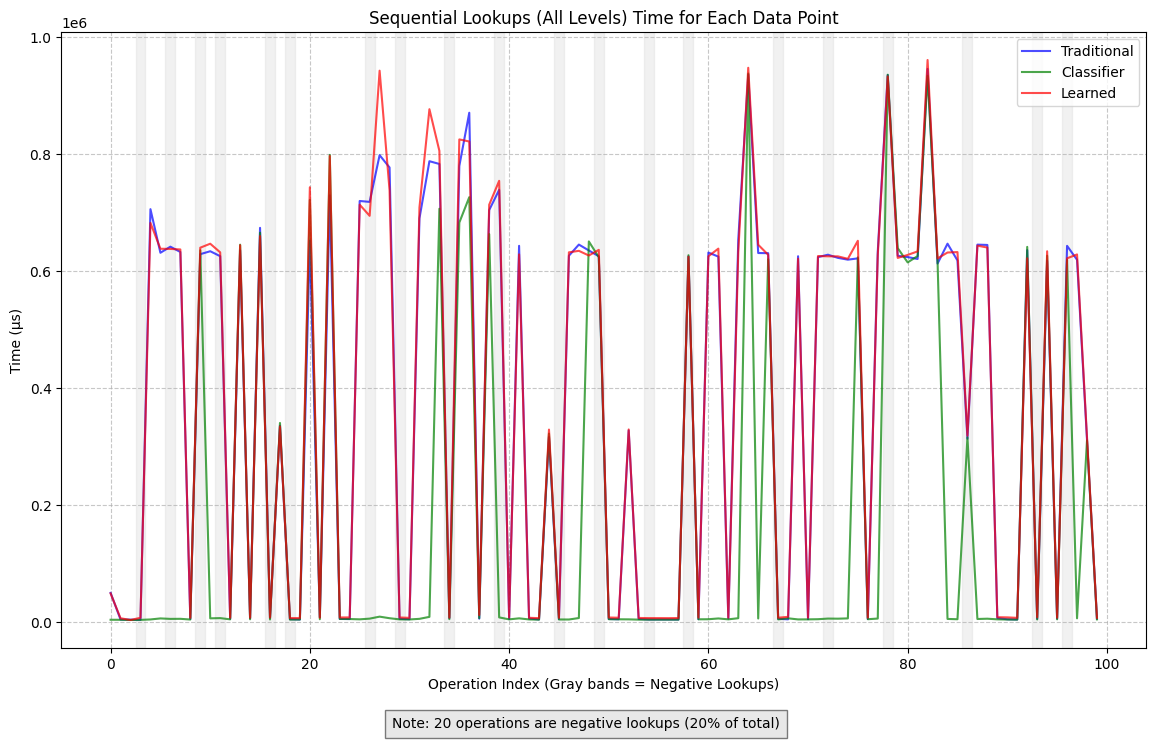}
    \caption{Sequential}
    \label{fig:detailed_sequential}
  \end{subfigure}

  \vspace{0.5em}

  \begin{subfigure}[b]{0.48\textwidth}
    \includegraphics[width=\textwidth]{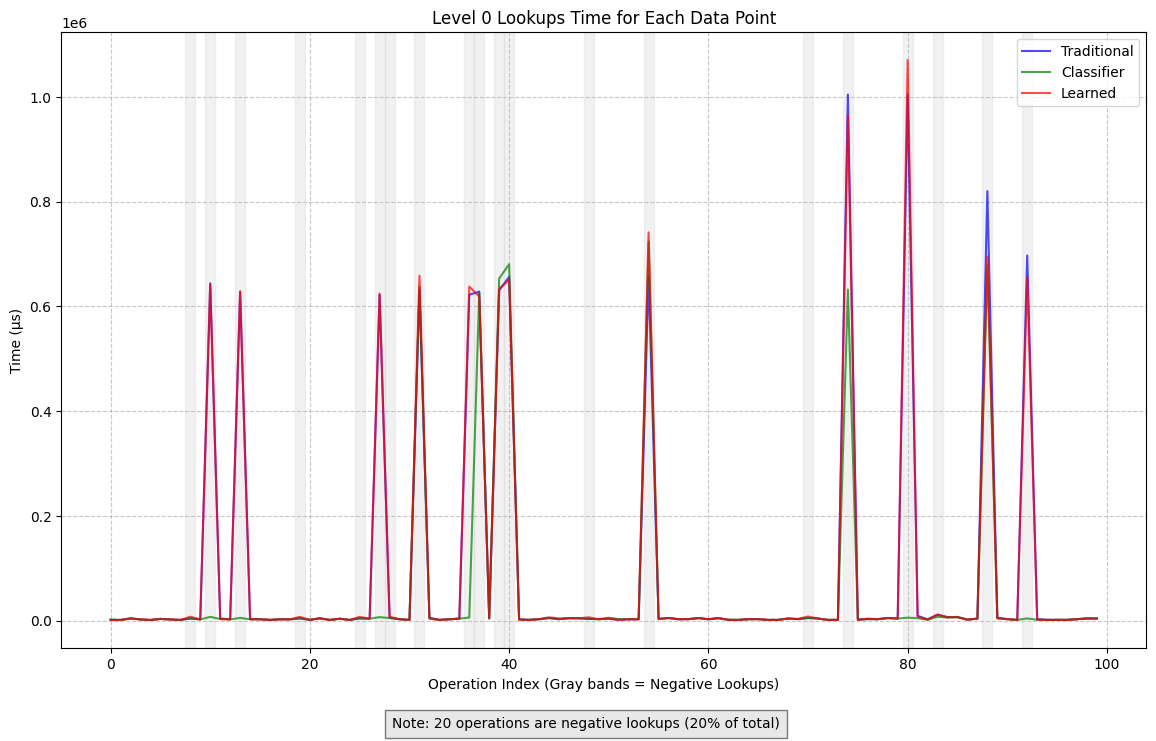}
    \caption{Level 0}
    \label{fig:detailed_l0}
  \end{subfigure}
  \hfill
  \begin{subfigure}[b]{0.48\textwidth}
    \includegraphics[width=\textwidth]{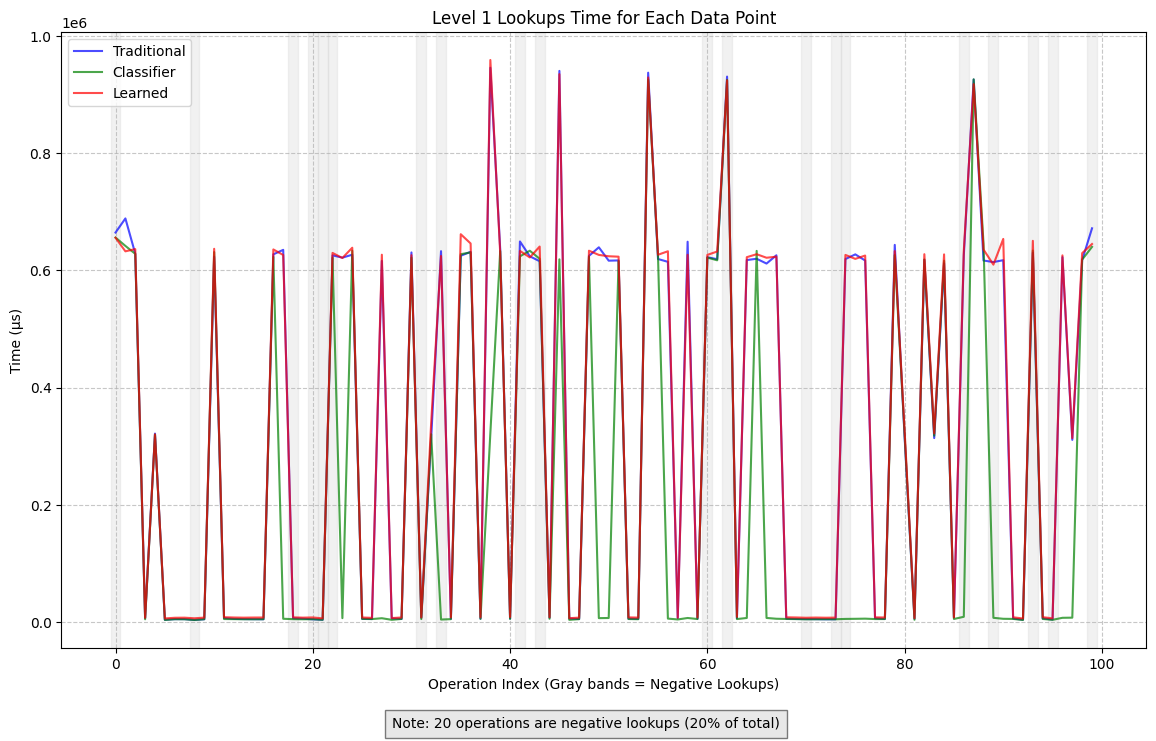}
    \caption{Level 1}
    \label{fig:detailed_l1}
  \end{subfigure}

  \vspace{0.5em}

  \begin{subfigure}[b]{0.48\textwidth}
    \includegraphics[width=\textwidth]{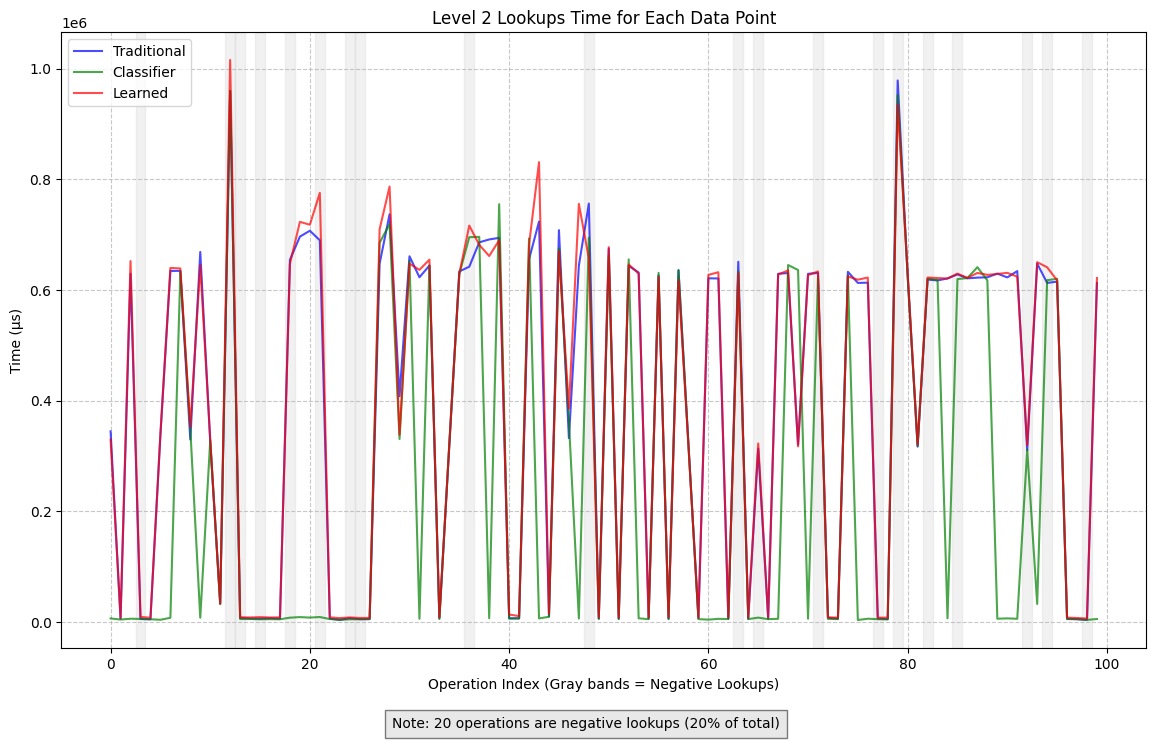}
    \caption{Level 2}
    \label{fig:detailed_l2}
  \end{subfigure}
  \hfill
  \begin{subfigure}[b]{0.48\textwidth}
    \includegraphics[width=\textwidth]{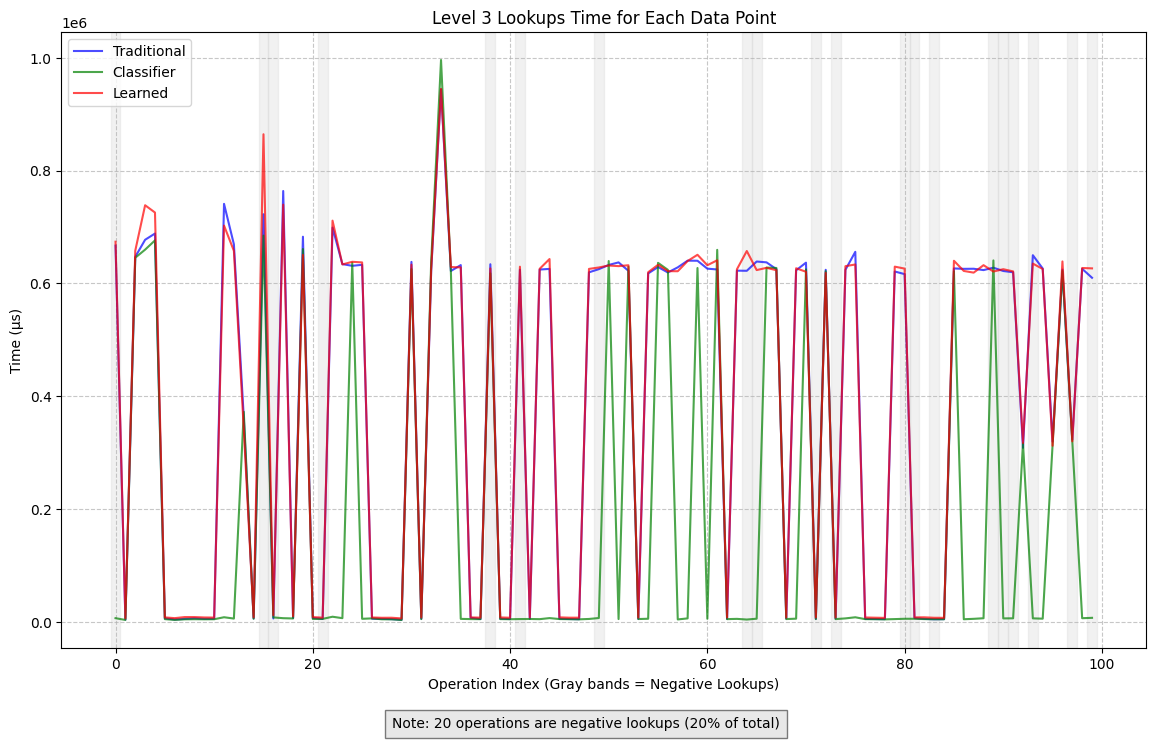}
    \caption{Level 3}
    \label{fig:detailed_l3}
  \end{subfigure}

  \caption{Per‐lookup latency profiles in a 3×2 grid.  
  Classifier dips indicate Bloom‐filter bypasses; Learned‐Bloom aligns with Traditional.}
  \label{fig:per_workload_detailed}
\end{figure}

\FloatBarrier
\clearpage

\begin{figure}[ht]
  \centering
  \includegraphics[width=0.85\linewidth]{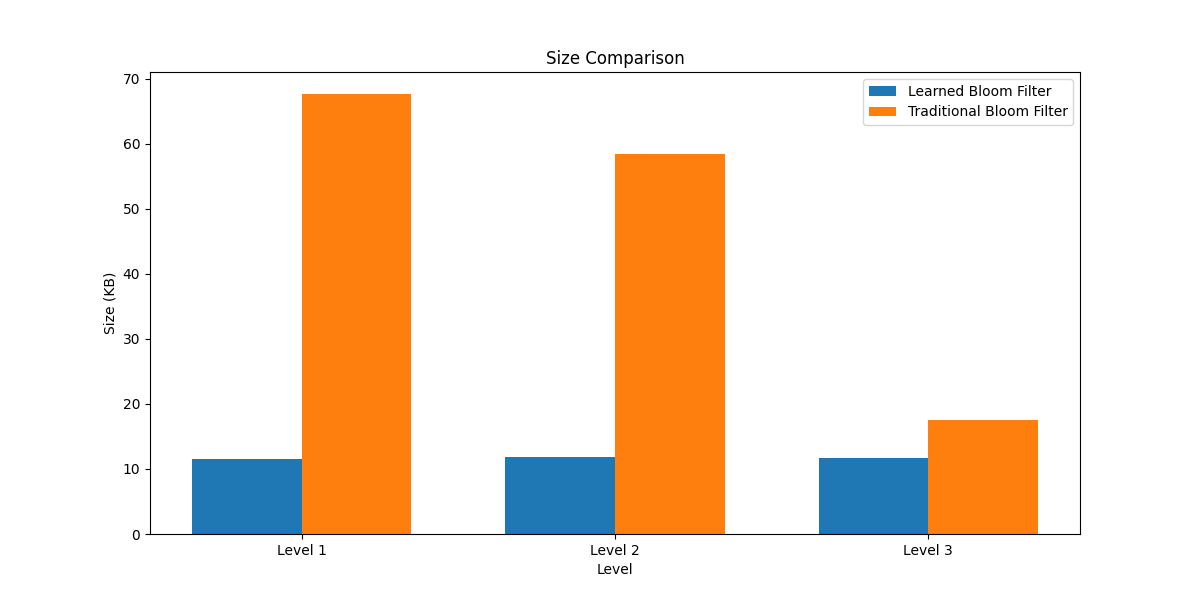}
  \caption{Memory usage comparison between traditional and learned Bloom filters across levels. Learned variant achieves substantial reductions in memory footprint.}
  \label{fig:learned_size_comparison}
\end{figure}

Figure~\ref{fig:learned_size_comparison} illustrates the Bloom filter memory footprint for levels 1 through 3, comparing traditional filters against their learned counterparts. At each level, the learned variant achieves a significant size reduction. For instance, at Level 1, the traditional filter consumes nearly 68 KB, while the learned Bloom filter occupies only 12 KB—a reduction of over 80\%. This pattern holds at Levels 2 and 3 as well, with savings consistent across the hierarchy. These gains validate the memory efficiency of hybrid learned filters, particularly when model inference latency remains within acceptable bounds.

\begin{figure}[ht]
  \centering
  \includegraphics[width=0.85\linewidth]{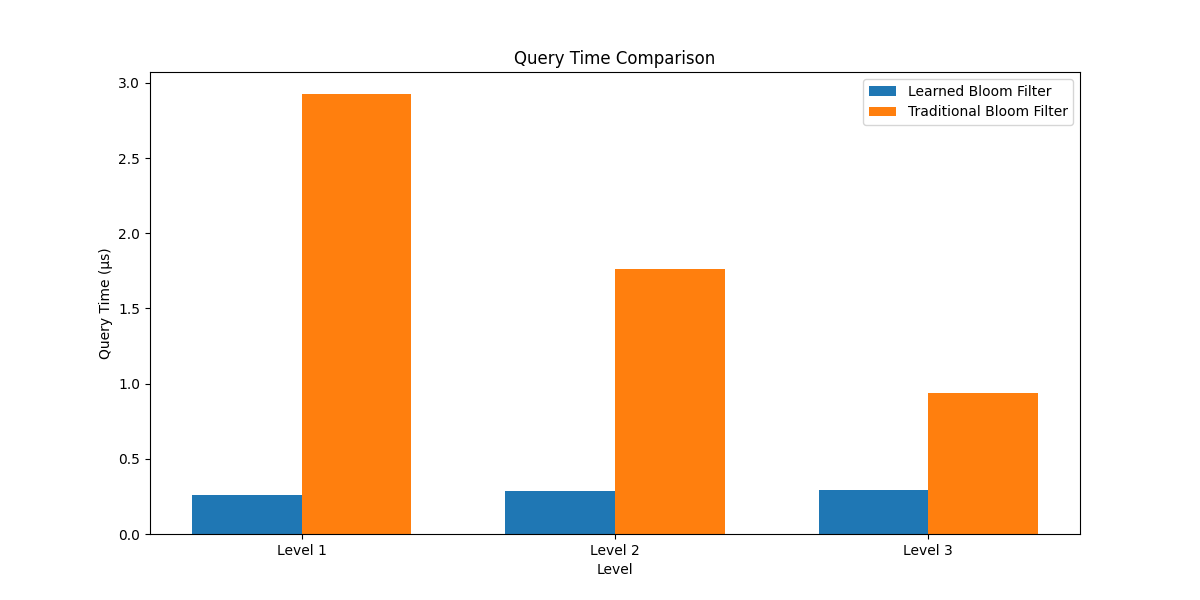}
  \caption{Average query time for traditional vs. learned Bloom filters across levels. Despite smaller size, learned variant retains competitive query times.}
  \label{fig:learned_time_comparison}
\end{figure}

Figure~\ref{fig:learned_time_comparison} complements the previous plot by comparing average query times between traditional and learned Bloom filters across the same three levels. Interestingly, the learned variant maintains performance parity despite the shift in internal structure. At Level 1, query latency drops from 2.9 $\mu$s (traditional) to under 0.3 $\mu$s (learned), indicating that compactness does not come at the cost of efficiency. Similar trends at Levels 2 and 3 reaffirm that learned filters can reduce both memory and access time simultaneously in certain settings.

\section{Discussion}

The experiments presented demonstrate two distinct pathways for integrating machine learning into the LSM‐tree read path: classifier‐guided filter skipping for latency reduction, and learned Bloom filters for memory efficiency without performance loss. Both approaches leverage predictive models to replace or augment traditional data structures, but they occupy different points in the design space and entail unique trade‐offs.

\subsection*{Latency vs.\ Memory Footprint}
At one extreme, the ML classifier approach optimizes pure lookup latency. By training a per‐level binary classifier on rich key features, we enable the system to avoid unnecessary Bloom‐filter probes on levels where the key is unlikely to reside. This yields dramatic speedups—up to 2.3× on cold Level 3 lookups—at the cost of introducing a nonzero false‐negative rate (FNR). In our experiments, the classifier skipped over 30–33\% of filters on average, translating directly into I/O savings and lower median latencies. However, the model itself occupies 5.6–6.1MB, and its rich feature set (45 engineered features such as powers, logs, trigonometric and digit statistics) contributes to both memory and computation overhead during inference.  

Conversely, the learned Bloom filter approach maintains strict zero‐FNR correctness while matching the traditional design’s performance almost exactly (within 1\%). By replacing large per‐level Bloom filters (which can consume multiple megabytes) with compact classifiers (530–903 KB) plus small backup filters, we reduce memory footprint by an order of magnitude without measurable latency penalty. This makes learned Bloom filters particularly attractive for memory‐constrained deployments—edge devices, mobile systems, or massive multi‐tenant clusters—where memory per key is at a premium.

\subsection*{False Negatives and Spike Analysis}
A detailed look at the classifier’s per‐operation latency profiles reveals that the number of downward “bypass” spikes far exceeds the count of actual false negatives reported by the model. This indicates that most speed gains arise not from mispredictions but from correctly skipping filter checks where the key genuinely does not exist. In other words, the classifier captures meaningful distributional patterns in the data—hot prefixes, common key ranges, or structural clustering—rather than merely exploiting its own errors. This insight suggests that even models with moderate FNR (e.g., 10–20\%) can achieve substantial latency benefits so long as their true‐negative precision is high.

\subsection*{Model Training and Deployment}
A critical concern for learned LSM‐trees is the cost and complexity of training and updating models in a live system. Our flush‐triggered training pipeline mitigates these concerns by decoupling model updates from query serving. Upon each MemTable flush, a background job locks only the metadata necessary to extract new key‐level labels, trains the classifier or learned Bloom filter on fresh data, and then atomically swaps in the new model once training completes. This nonblocking, versioned approach ensures continuous availability with no service interruption.  Moreover, incremental training techniques—warm‐starting from previous model weights, streaming feature updates, and online gradient updates—can further reduce training latency and enable true real‐time adaptability to workload shifts.

\subsection*{Scalability and Sampling Strategies}
As database size grows, so do model training costs and feature engineering challenges—especially for lower levels containing millions of keys. In our implementation, we subsampled 10–20\% of Level 2 and Level 3 keys to bound training time. While effective, this sampling can degrade model accuracy on rarely accessed tail keys, contributing to the slightly higher FNR observed at deeper levels. Future work should explore adaptive sampling—over‐sampling rare key ranges, under‐sampling dense hot ranges, or using stratified sampling based on access frequency—to balance representativeness with training cost. Alternatively, online learning algorithms that continuously adjust model parameters on individual operations could eliminate the need for explicit sampling altogether.

\subsection*{Dynamic Thresholding and Model Calibration}
Our current classifiers use a fixed decision threshold (0.5) to determine filter skipping. However, different workloads and levels may benefit from customized thresholds—lowering the bar in deeper levels to reduce false negatives, or raising it in hot levels to maximize bypass rates. Implementing dynamic thresholding, based on real‐time monitoring of FPR/FNR trade‐offs, could yield additional gains.  Similarly, calibrating model confidence (e.g., via temperature scaling) would allow the system to adjust its aggressiveness in response to workload changes, striking a balance between latency, accuracy, and memory usage.

\subsection*{Predictive Fence Pointers and Multiclass Models}
Beyond Bloom filters, fence pointers are another natural candidate for learned acceleration. In a standard LSM-tree, fence pointers divide each SSTable into fixed‐size page ranges and use binary search over these pointers to locate a key—an \(O(\log P)\) operation where \(P\) is the number of pages. A small regression model could instead predict the page index directly, yielding an \(O(1)\) jump to the approximate offset. Because binary search over tens or hundreds of pages is already quite fast, the absolute latency reduction may be modest, but even a few cache‐line savings per lookup can compound across high‐throughput workloads. Implementing and evaluating such a model would clarify whether learned fence pointers can meaningfully supplement or replace pointer arrays in production systems.

Our current classifier design uses independent binary models at each level. An alternative is a single multiclass predictor that directly outputs the most likely level for a given key. This could eliminate multiple per‐level inferences and simplify the lookup path. Initial experiments with multiclass forests and shallow neural nets yielded lower accuracy—likely due to the larger label space—but suggest that more expressive architectures (e.g., boosted trees with level‐aware features or transformer‐based key embeddings) might succeed. A reliable multiclass model could also drive prefetching or cache warming, loading the predicted SSTable’s pages ahead of the actual lookup and further reducing end‐to‐end latency.

\subsection*{Co‐Design with Compaction and Layout Policies}
Learned models do not exist in isolation—their efficacy depends on the underlying data layout and compaction policy. For instance, a more aggressive leveling schedule that reduces SSTable overlap could simplify model features and improve prediction accuracy. Conversely, a tiered compaction policy that retains more cold data in L0 might favor learned Bloom filters by reducing model complexity for deeper levels. Jointly tuning compaction thresholds, level size ratios, and model architectures in an end‐to‐end co‐design framework is an exciting direction for future research.

\subsection*{Workload Adaptivity and Continuous Learning}
Finally, real‐world access patterns evolve over time. The ability to continuously monitor model performance—tracking drift in FNR, bursts in false positives, or shifts in bypass efficacy—and to trigger retraining or model reconfiguration is essential for long‐lived systems. Building an adaptive control loop that integrates telemetry, automatic retraining, and dynamic kernel updates would close the gap between research prototypes and production‐ready learned LSM‐trees.

\bigskip

In summary, our work demonstrates the feasibility and benefits of learned auxiliary structures in LSM‐trees, and opens multiple avenues for refinement: from adaptive sampling and threshold tuning, through predictive fence pointers and multiclass predictors, to co‐design with compaction policies and continuous learning frameworks.  These insights underscore the rich potential of blending data systems with modern learning techniques.

\bibliographystyle{plain}
\bibliography{references}

\end{document}